\documentclass[3p,preprint,12pt]{elsarticle}
\usepackage{graphics}
\usepackage{ecrc}
\usepackage{amssymb}
\usepackage{epsfig}

\usepackage{adjustbox}

\volume{00}

\firstpage{1}

\journalname{Nuclear Physics A}

\runauth{}

\jid{procs}

\jnltitlelogo{Nuclear Physics A}

\journal{Nuclear Physics A}

\begin{document}

\begin{frontmatter}

\title{High precision half-life measurement of $^{125}$Cs and $^{125}$Xe with $\gamma$-spectroscopy}

\author[ATOMKI]{T. N. Szegedi}
\author[ATOMKI]{G. G. Kiss \corref{cor}}
\ead{ggkiss@atomki.mta.hu}
\cortext[cor]{corresponding author}
\author[KOCA]{I. \"Oks\"uz \fnref{fn1}}
\fntext[fn1]{Present Address: Department of Mechanical and Aerospace Engineering, 
The Ohio State University, 43210 OH Columbus, USA}
\author[ATOMKI]{T. Sz\"ucs \fnref{fn2}}
\fntext[fn2]{Present Address: Helmholtz-Zentrum Dresden-Rossendorf
(HZDR), 01328 Dresden, Germany}
\author[ATOMKI]{Gy. Gy\"urky}
\author[ATOMKI]{Z. Elekes}
\author[ATOMKI]{E. Somorjai}
\author[ATOMKI]{Zs. F\"ul\"op}

\address[ATOMKI]{Institute for Nuclear Research (MTA Atomki), H-4001 Debrecen, Hungary}
\address[DE]{University of Debrecen, H-4001 Debrecen, Hungary}
\address[KOCA]{Kocaeli University, TR-41380 Umuttepe, Kocaeli, Turkey}

\begin{abstract}
In order to provide data for the simulation of the astrophysical $\gamma$-process, the cross section measurement of the $^{124}$Xe(p,$\gamma$)$^{125}$Cs reaction is in progress at MTA Atomki using the activation technique. Precise information on the decay characteristics of the reaction products is of crucial importance for measurements carried out using this method. The half-lives of the produced $^{125}$Cs and its daughter $^{125}$Xe are published in previous works, but with large uncertainties and ambiguous values. To make these nuclear parameters more precise, the half-lives have been re-measured with high precision using $\gamma$-spectroscopy. The obtained new half-life values are t$_{1/2}$ = 44.35 $\pm$ 0.29 minutes for $^{125}$Cs and t$_{1/2}$ = 16.87 $\pm$ 0.08 hours for $^{125}$Xe. 
\end{abstract}


\end{frontmatter}

\section{Introduction}
\label{sec:intro}

The stable, heavy proton-rich -- so\,-\,called $p$ -- isotopes are produced in explosive nucleosynthesis scenarios \cite{Rau13}. The main stellar mechanism creating these species, the so\,-\,called $\gamma$-process, involves photodisintegration reactions on the available s and r seed nuclei. It is generally accepted that the $\gamma$-process takes place either in the O/Ne-rich layer of type II supernovae during core collapse or during the thermonuclear explosion of white dwarfs \cite{Rau02, Tra11}. Regardless of the astrophysical site, the modeling of the $\gamma$-process requires the use of extended nuclear reaction network calculations. Due to the lack of experimental data, the reaction rates are mainly provided by the Hauser\,-\,Feshbach statistical model. For the model calculations, nuclear physics inputs -- e.g. optical model potentials, nuclear level densities, $\gamma$-ray strength functions, etc. -- are needed and the ambiguities of these input parameters introduce considerable uncertainties to the reaction rate predictions and consequently to the calculated $p$ isotope abundances \cite{Rap06}.

Therefore, the measurement of the relevant reaction cross sections is of high importance to check the predictions of the statistical model calculations and to select the proper input parameter set. From sensitivity studies it is known that below the A $\approx$ 140 region the ($\gamma$,p) and above the ($\gamma$,$\alpha$) reactions are particularly important \cite{Rap06}. Due to experimental difficulties \cite{Moh07, Kis08, Rau09}, instead of the photodisintegrations the inverse capture reactions should be studied experimentally. For this purpose, in recent years proton and alpha capture reactions were studied extensively (see \cite{Gyu14, Net14, Kis14, Sch14} and references therein), however, the available database is still not sufficient \cite{kadonis}. The absence of experimental cross section values for proton-induced reactions on $^{124}$Xe led us to perform a measurement. We used the well-known activation technique where the cross section is deduced from the off-line activity measurement of the reaction products, $^{125}$Cs and its daughter $^{125}$Xe.

\subsection{Available half-life values}

The decay parameters of the reaction products enter into the calculations, consequently their precise values must be known and their uncertainty contributes to the uncertainty of the obtained experimental cross sections. About 50 years ago three experiments were devoted to measure the half-life of $^{125}$Cs \cite{Mor54, Pre62, Aur67}. The values were determined in these works with relatively large uncertainties (2.2\%, 10.2\% and 8.7\%, respectively) and furthermore, the data of \cite{Aur67} is available only in an unpublished thesis and the work of \cite{Mor54} provides only very limited experimental information. In the compilations before 1981 \cite{Tam81} the weighted average (45.2 $\pm$ 1 min) of the half-lives was given as the recommended value. However, this was changed -- without any detailed explanation -- in the compilations issued after 1993 
\cite{Kat11} and the unweighted average of the three measurements \cite{Mor54, Pre62, Aur67} (46.7 min) was given. The uncertainty assigned to this unweighted value (0.1 min)\cite{Kat11} is obviously a typo.

\begin{center}
\begin{table}
\caption{\label{tab:T1/2} Half-lives of the $^{125}$Cs and $^{125}$Xe isotopes taken from the literature.} 
\begin{tabular}{clcc}
\hline 
\multicolumn{1}{c}{\rule{0pt}{2.5ex} Isotope} &
\multicolumn{1}{c}{\rule{0pt}{2.5ex} Half-life} &
\multicolumn{1}{c}{\rule{0pt}{2.5ex} Year of the pub.} &
\multicolumn{1}{c}{\rule{0pt}{2.5ex} Reference} \\
\hline
\rule{0pt}{2.5ex}
$^{125}$Cs  &  45.0  $\pm$ 1.0 min  & {\it 1954}  & \cite{Mor54} \\
            &  49.0  $\pm$ 5.0 min  & {\it 1962}  & \cite{Pre62} \\
            &  46.0  $\pm$ 4.0 min  & {\it 1967}  & \cite{Aur67} \\
recommended value (unweighted average) & 46.7 $\pm$ 0.1 min\footnotemark & & \cite{Kat11}  \\
weighted average & 45.2 $\pm$ 1 min & & \\ 
\hline
\rule{0pt}{2.5ex}  
$^{125}$Xe  &  20.0  $\pm$ 1.0 h    & {\it 1950} & \cite{And50} \\
			      &  18.0  $\pm$ 0.4 h  & {\it 1952} & \cite{Ber52} \\
            &  17.0  $\pm$ 1.0 h    & {\it 1960} & \cite{Moo60} \\	
			      &  16.8  $\pm$ 0.2 h  & {\it 1965} & \cite{Gan65} \\
            &  17.3  $\pm$ 0.4 h  & {\it 1969} & \cite{Lud69} \\
recommended value (weighted average of \cite{Gan65,Lud69} & 16.9 $\pm$ 0.2 h & & \cite{Kat11}\\
weighted average (all data) & 17.2 $\pm$ 0.3 h\footnotemark & & \\
\hline
\end{tabular}
\end{table}
\end{center}

The situation is somewhat different in the case of $^{125}$Xe: although more (the half-life data is based on five measurements \cite{And50, Ber52, Moo60, Gan65, Lud69}) and better documented data is available, the p-value of the weighted mean is only 0.3\% which indicates that the systematic uncertainties of one or more measurements  might be underestimated. Moreover, similarly to the $^{125}$Cs case, the relative uncertainties are large (5.0\%, 2.2\%, 6.0\%, 2.0\% and 2.3\%, respectively) and in these works there are very limited discussions about the experimental setup and on data analysis. In the first compilation issued in 1972 \cite{Aub72} the unweighted average (17.0 $\pm$ 0.3 h) of the results presented in \cite{Gan65, Lud69} was given as the recommended value. This value was changed in 1981: the weighted average of \cite{Gan65, Lud69} (16.9 $\pm$ 0.2 h) is indicated as the recommended value in the more recent compilations 
\cite{Kat11}. The above discussed $^{125}$Cs and $^{125}$Xe half-life data are listed in Table \ref{tab:T1/2}.

In summary, an independent confirmation of the half-lives or providing more reliable data is of crucial importance for our forthcoming $^{124}$Xe(p,$\gamma$)$^{125}$Cs activation measurements. The aim of this paper is to present our experimental procedure to provide more precise half-life values. This paper is organized as follows: after a short introduction to our motivation in Sect. 1, in Sect. 2 the technical details of the experimental procedure --- including the description of the source preparation and the $\gamma$-counting setup --- are presented and the data analysis is explained, while in Sect. 3 the results are given.

\footnotetext[3]{The assigned uncertainty is obviously a typo.}
\footnotetext{The uncertainty is scaled by the square root of the reduced chi square.}


\begin{center}
\begin{figure}
\resizebox{1.0\columnwidth}{!}{\rotatebox{0}{\includegraphics[clip=]{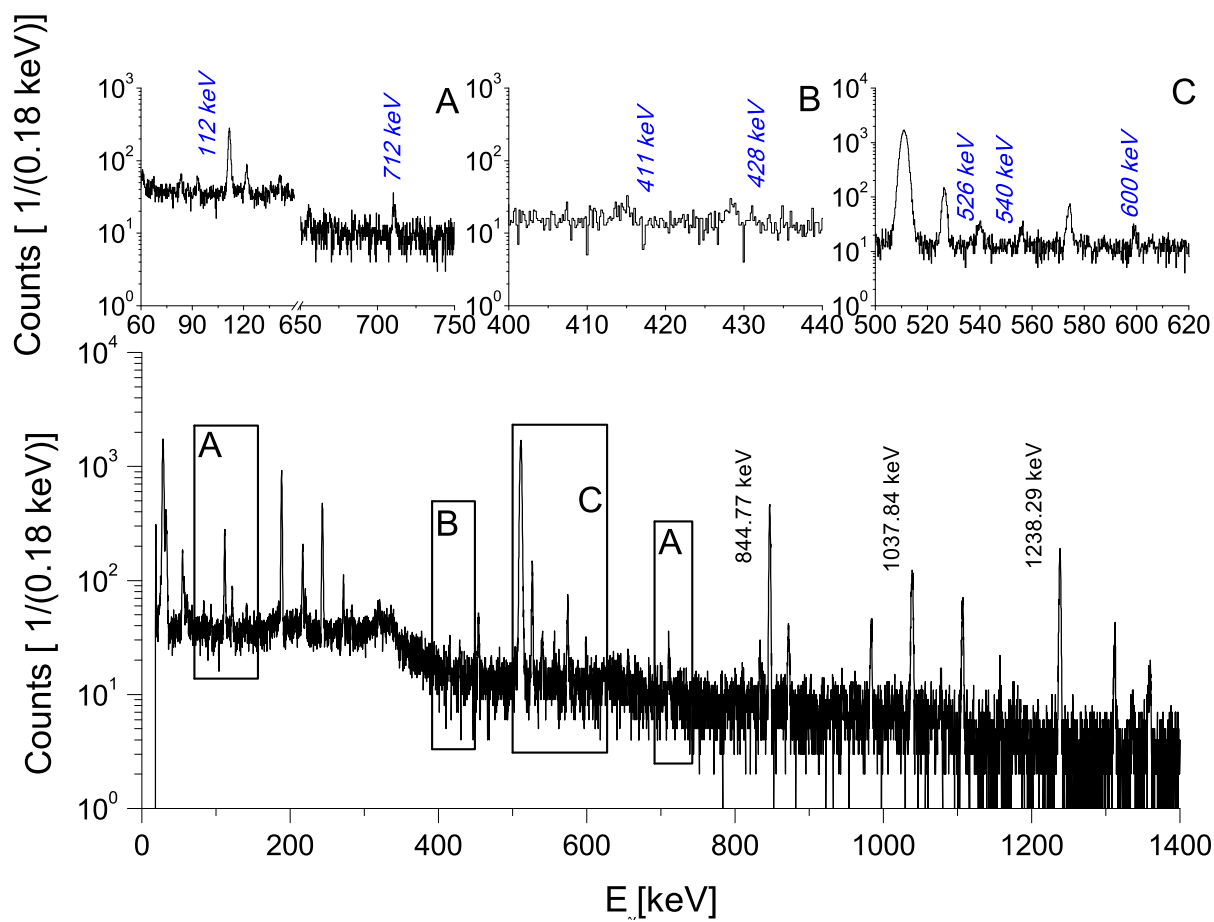}}}
\caption{\label{fig:125Cs_spectrum} A 3 min $\gamma$-spectrum measured 67 min after the end of the irradiation by an E$_p$ = 7.4\,MeV proton beam. The insets (A,B,C) show the relevant energy regions with the most important peaks of $^{125}$Cs (blue, italic). A $^{56}$Co source was used to test the stability of the $\gamma$-counting setup, its peaks are also indicated (black).}
\end{figure}
\end{center}

\section{Experimental procedure}
\label{sec:exp_proc}


\subsection{Target preparation and irradiation}
\label{sec:source}

Recently, a thin window gas-target system \cite{Bor12} was designed at Atomki in order to measure low energy $\alpha$-induced reaction cross sections relevant for the $\gamma$-process and for measuring the $^3$He($\alpha$,$\gamma$)$^7$Be reaction using the activation method \cite{Hal16, Bor13}. This chamber was slightly modified according to the special experimental requirements of the $^{124}$Xe(p,$\gamma$)$^{125}$Cs cross section measurement. Namely, the length of the chamber was reduced to 4.2 mm, for the entrance window a 10.2 $\mu$m thick aluminum foil was used and the $^{125}$Cs recoils were implanted into a similar high purity aluminum (catcher-) foil.  
The $^{125}$Cs sources were produced by the $^{124}$Xe(p,$\gamma$)$^{125}$Cs reaction, the proton beam was provided by the MGC-20 cyclotron accelerator of Atomki. Six sources were produced with proton bombarding energies of 6.0 MeV, 6.5 MeV (carried out twice), 7.0 MeV, 7.4 MeV and 7.5 MeV. In this energy region the proton capture cross section reaches its maximum values according to the statistical model calculations carried out by the NON-SMOKER code \cite{rau_web}. The isotopic purity of the $^{124}$Xe gas was better than 99.9\% and the lengths of the irradiation varied between 70 and 180 minutes. 
Between the end of the irradiation and the beginning of the $\gamma$-counting 30-60 minutes elapsed.

\subsection{The counting setup and its validation}
\label{sec:gamma}
  
The half-lives of the $^{125}$Cs and $^{125}$Xe isotopes were measured with $\gamma$-spectroscopy. After the irradiations the catcher foil was removed from the gas chamber and transported to the off-line counting setup where the $\gamma$-radiation following the $\beta$ decay of the reaction products was measured using a 100\% relative efficiency HPGe detector (Canberra model GR10024), placed into a complete 4$\pi$ multilayered shielding\footnotemark. This setup was also used in the past for the precise half-life measurements carried out at Atomki \cite{Gyu09, Far11, Gyu12}. A similar data acquisition system to \cite{Gyu09} was used for this experiment. Namely, the signals from the detector preamplifier were first shaped (the shaping time was set to 3 $\mu$s) and amplified by an ORTEC model 671 module. An ORTEC model ASPEC-927 multichannel analyzer unit was used and the data were collected using the ORTEC A65-B32 MAESTRO software which provides deadtime information found to be precise in the previous works \cite{Gyu09, Gyu12}. The energy calibration was carried out using calibrated $^{60}$Co, $^{133}$Ba, $^{137}$Cs, $^{152}$Eu and $^{241}$Am sources (for more details see e.g. \cite{hal12}). 

\footnotetext{The type of the shielding, purchased from Tema Sinergie S.P.A (https://www.temasinergie.com/), is Tema Sinergie Mod. GDS 2.}

\begin{center}
\begin{figure}
\resizebox{1.0\columnwidth}{!}{\rotatebox{0}{\includegraphics[clip=]{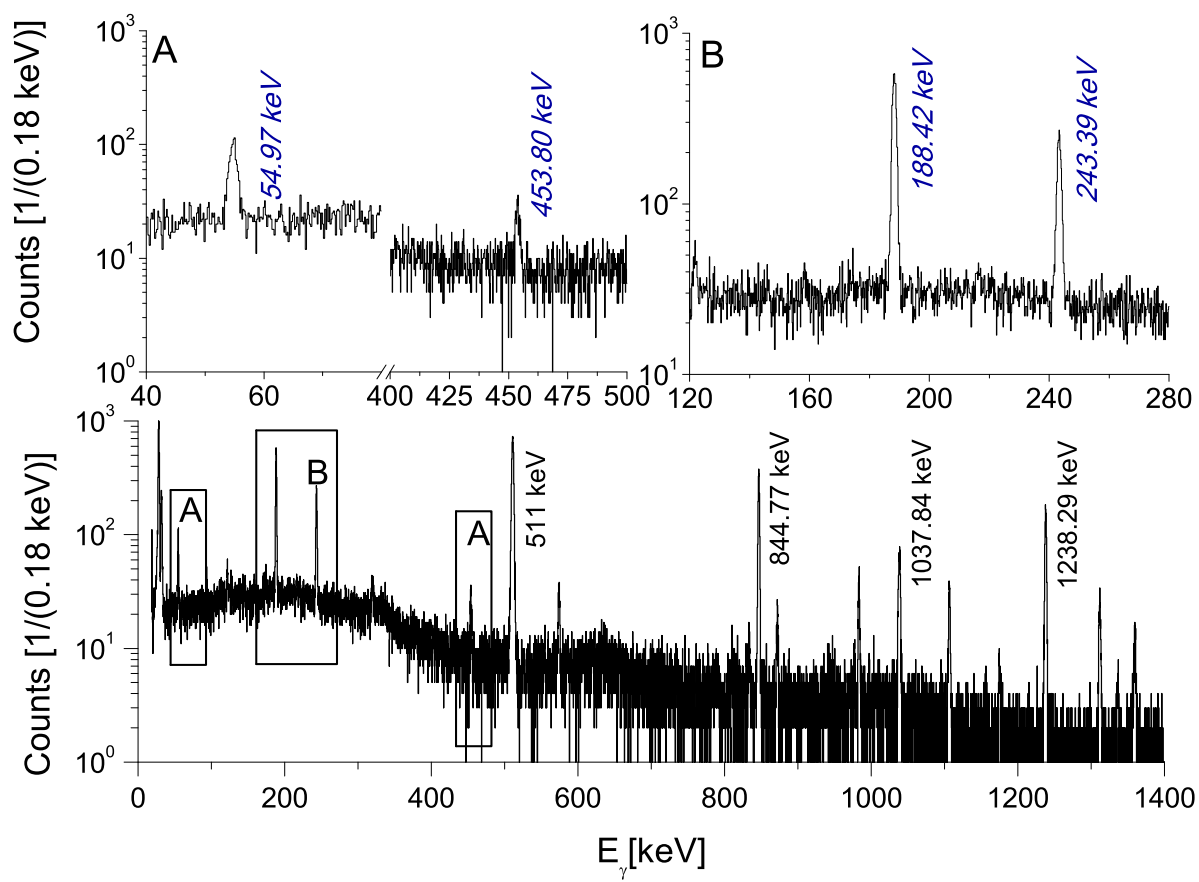}}}
\caption{\label{fig:125Xe_spectrum} A 0.25 hour $\gamma$-spectrum measured 15.1 hours after the end of the E$_p$ = 7.4 MeV irradiation. The insets (A,B) show the relevant energy regions with the most important peaks of $^{125}$Xe (blue, italic) and the $^{56}$Co peaks (black) used to monitor the stability of the counting setup.}
\end{figure}
\end{center}

\subsection{Determination of $^{125}$Cs and $^{125}$Xe half-lives}

The activity of each irradiated sample was measured for 29 to 61 hours\footnote{Except for the counting of the 2nd source produced at E$_p$ = 6.5 MeV which had to be stopped due to technical problems.}. In the first 5 hours of the counting the spectra were saved and then erased in every 3 minutes to derive the half-life of the $^{125}$Cs isotope and later in every 15 minutes to measure the half-life of the $^{125}$Xe nucleus. For each spectra the deadtime value, used to correct the corresponding peak area, was provided by the MAESTRO software and was found to be between 0.11\% and 2.19\%, depending on the time elapsed after the irradiation and on the counting geometry.  

Figures \ref{fig:125Cs_spectrum} and \ref{fig:125Xe_spectrum} shows a 3 min and a 15 min $\gamma$-spectrum, taken 67 minutes and 15.1 hours after the beginning of the $\gamma$-counting following the E$_p$ = 7.4 MeV proton beam irradiation. In table \ref{tab:counting} the details of the gamma countings are summarized. 

The sources produced by E$_p$ = 6\,MeV and by E$_p$ = 6.5\,MeV proton irradiation were placed at 1 cm distance from the end-cap of the detector. The activity measurements of the sources produced by E$_p$ = 7.0\,MeV, E$_p$ = 7.4\,MeV and E$_p$ = 7.5\,MeV irradiations were carried out using two different geometries: these samples were placed at first at 10 cm (far geometry) from the end-cap of the detector and the decay of the $^{125}$Cs isotope was measured for 1.8 - 2.15 hours. Thereafter, the sources were moved to 1 cm (close geometry) from the end-cap of the detector and the activity was measured again. In these cases the counting length was within 1.4 - 2.75 hours. 
To derive the half-life of the $^{125}$Xe isotope the activity was measured similarly first in far geometry and then in close geometry. The total lengths of the countings were about 27-38 hours at both geometries. 

Although both isotopes decay by the emission of several $\gamma$-rays \cite{Kat11}, unfortunately the relative intensities of most of the transitions were too low for a precise half-life determination. Furthermore, due to the limited resolution of the HPGe detector, it was not possible to resolve the E$_{\gamma}$ = 112 keV $\gamma$-ray in $^{125}$Cs from the E$_{\gamma}$ = 113.55 keV $\gamma$-ray in $^{125}$Xe. Since the subtraction of the parasitic contribution would lead to noticeably higher uncertainties we omit this transition from the half-life determination. Therefore, to derive the half-life of the $^{125}$Cs isotope the area of the 526 keV peak was determined in each spectra measured for 3 min  
The determination of the $^{125}$Xe half-life is based on two $\gamma$-lines, the rates of the E$_{\gamma}$ = 188.42 keV and E$_{\gamma}$ = 243.38 keV $\gamma$-transitions were sufficient in each 15 min long counting for a precise measurement. 

\begin{center}
\begin{figure*}
\resizebox{1.0\columnwidth}{!}{\rotatebox{0}{\includegraphics[clip=]{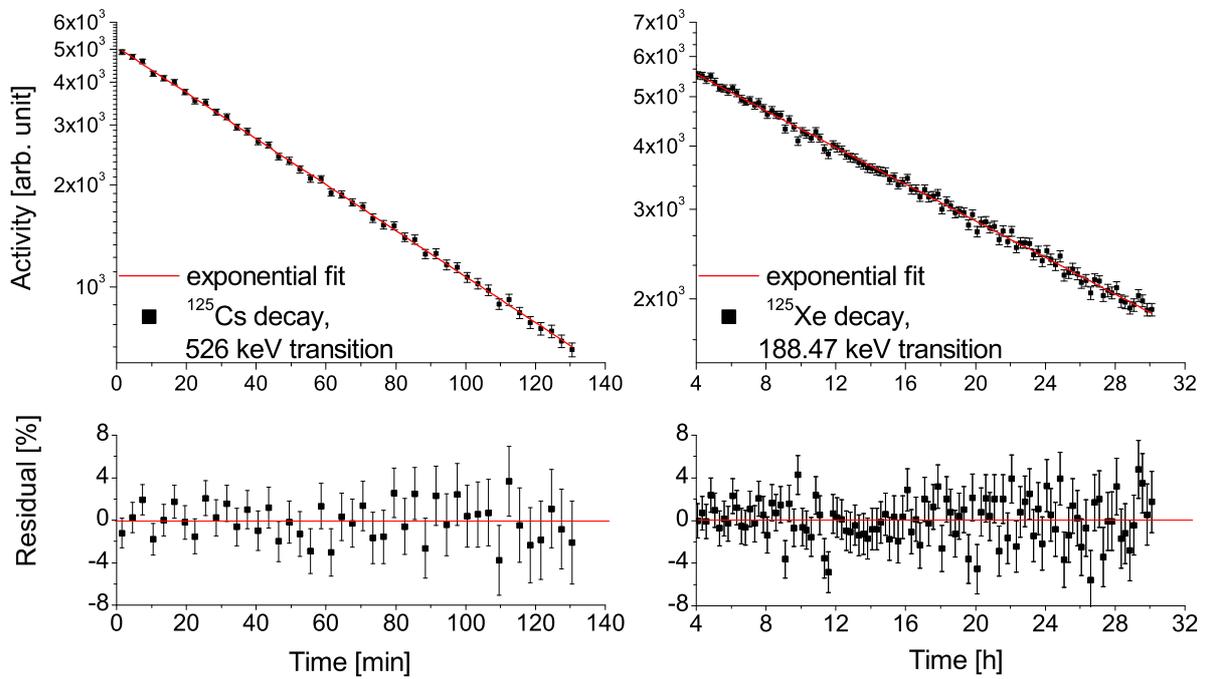}}}
\caption{\label{fig:decay_curve} Decay curves (and residuals) based on the study of the E$_{\gamma}$ = 526 keV ($^{125}$Cs) and E$_{\gamma}$ =188.42 keV ($^{125}$Xe) $\gamma$-transitions. The widths of each time bin are 3 min for $^{125}$Cs and 15 min for $^{125}$Xe, respectively. 
}
\end{figure*}
\end{center}

Two methods --- resulting in half-lives always within 0.2\% --- were used to determine the net peak areas. In each spectra the peaks were fitted with a Gaussian plus a linear background and the peak areas were also determined by numerical integration. In the two cases different background regions were selected. As can be seen in figs. \ref{fig:125Cs_spectrum} and \ref{fig:125Xe_spectrum} there are parts on both sides of the region of interest where the background for the peak area determination could be fixed with good confidence. A linear function was fitted onto the two background regions and the background was subtracted from the peak areas. The determination of the half-lives was made by fitting an exponential curve --- using the least square method (described in details e.g. in \cite{Leo92}) --- on the deadtime corrected areas of the $\gamma$-peaks as the function of the elapsed time. Fig. \ref{fig:decay_curve} shows the decay curve of $^{125}$Cs (left side) and $^{125}$Xe (right side) with the measured points and the fitted exponential function as well as the percentage residuals (bottom). 

The rates of the weaker transitions, corresponding to the decay of $^{125}$Cs and $^{125}$Xe, in each 3 min and 15 min long spectra are about factor of 5 to 10 lower. Several spectra were summed up and the rate of the weaker transitions were studied. Although we found that the t$_{1/2}$ values based on the counting of these transitions scatter around the t$_{1/2}$ value proclaimed here, since their statistical uncertainties are significantly larger (about 2.2 \% - 5.1 \%), involving these transitions in the weighted mean does not improve the precision of the final result.

\begin{center}
\begin{table}
\caption{\label{tab:counting} Details of the $\gamma$-counting. The energy of the proton beam, the geometry of the $\gamma$-counting and the corresponding counting length are indicated. Furthermore, the initial total counting rate characterizing the first spectra of each run and the corresponding maximum deadtime values are listed. After the E$_p$ = 7 MeV, E$_p$ = 7.4 MeV and E$_p$ = 7.5 MeV irradiations the source activities were measured at first in far (far$\#$1) then in close (close$\#$1) geometry, in total for about 3.5-5 hours to obtain the half-life of the $^{125}$Cs isotope. Similarly after these irradiations, the counting used to derive the t$_{1/2}$ of the $^{125}$Xe were performed using different source-to-detector distance, these countings are marked with far$\#$2 and close$\#$2.} 
\begin{tabular}{cccccc}
\hline 
\multicolumn{1}{c}{\rule{0pt}{2.5ex} E$_p$} &
\multicolumn{1}{c}{\rule{0pt}{2.5ex} Counting} &
\multicolumn{1}{c}{\rule{0pt}{2.5ex} Total counting} &
\multicolumn{1}{c}{\rule{0pt}{2.5ex} Initial total } &
\multicolumn{1}{c}{\rule{0pt}{2.5ex} Initial} \\
\multicolumn{1}{c}{\rule{0pt}{2.5ex} [MeV]} &
\multicolumn{1}{c}{\rule{0pt}{2.5ex} geometry } &
\multicolumn{1}{c}{\rule{0pt}{2.5ex} length [hour]} &
\multicolumn{1}{c}{\rule{0pt}{2.5ex} count rate [Hz]} &
\multicolumn{1}{c}{\rule{0pt}{2.5ex} deadtime [\%]} \\
\hline
\rule{0pt}{2.5ex}
6   & close & 61.3 & 1.32 x 10$^3$& 2.19 \\
6.5 & close & 9.55  & 1.33 x 10$^3$& 2.17 \\  
6.5 & close & 2.8 & 1.19 x 10$^3$ & 2.14 \\
7.0 & far$\#$1   & 2.15 & 6.23 x 10$^2$  & 1.04 \\
 & close$\#$1 & 2.75 & 5.44 x 10$^2$  & 0.92 \\
 & far$\#$2   & 37.5 & 6.31 x 10$^1$& 0.11 \\
 & close$\#$2 & 28.75& 1.49 x 10$^2$& 0.28 \\
7.4 & far$\#$1   & 1.8  & 8.27 x 10$^2$ & 1.57 \\
 & close$\#$1 & 2.1  & 8.02 x 10$^2$ & 1.46 \\
 & far$\#$2   & 26.75& 2.24 x 10$^2$& 0.44 \\
 & close$\#$2 & 35   & 4.67 x 10$^2$& 0.81 \\
7.5 & far$\#$1   & 2.15 & 1.29 x 10$^3$& 2.02 \\
 & close$\#$1 & 1.4  & 1.28 x 10$^3$& 2.22 \\
 & far$\#$2   & 36.25& 2.08 x 10$^2$& 0.33 \\
 & close$\#$2 & 33.75& 5.15 x 10$^2$& 0.95 \\
\hline
\end{tabular}
\end{table}
\end{center}

\begin{center}
\begin{table}
\caption{\label{tab:res} Half-life results. The listed uncertainties are statistical only. The countings marked with $^*$ were carried out in the far geometry. For more details see text.}
\begin{tabular}{cccccc}
\hline
\multicolumn{1}{c}{\rule{0pt}{2.5ex}  Isotope} &
\multicolumn{1}{c}{\rule{0pt}{2.5ex}  E$_{\mbox {\small proton}}$ [MeV]} &
\multicolumn{1}{c}{\rule{0pt}{2.5ex}  E$_{\gamma}$ [keV]} &
\multicolumn{1}{c}{\rule{0pt}{2.5ex}  t$_{1/2}$ [min]} &
\multicolumn{1}{c}{\rule{0pt}{2.5ex}  d.o.f.} &
\multicolumn{1}{c}{\rule{0pt}{2.5ex}  $\chi^{2}_{red.}$} \\
\hline
\rule{0pt}{2.5ex}  
$^{125}$Cs& 6.0 & 526 & 44.63$\pm$0.20 & 56 & 1.10 \\
          & 6.5 & 526 & 44.36$\pm$0.16 & 56 & 0.92 \\
          & 6.5 & 526 & 44.34$\pm$0.19 & 56 & 1.01 \\
					& 7.0$^*$ & 526 & 44.02$\pm$0.73 & 43 & 1.07 \\
					& 7.0 & 526 & 44.32$\pm$0.29 & 55 & 0.99 \\
					& 7.4$^*$ & 526 & 44.16$\pm$0.63 & 36 & 1.37 \\
					& 7.4 & 526 & 44.01$\pm$0.19 & 42 & 1.02 \\
					& 7.5$^*$ & 526 & 44.59$\pm$0.41 & 43 & 0.96\\
					& 7.5 & 526 & 44.46$\pm$0.27 & 28 & 1.26 \\			
Error weighted mean&  & & 44.35$\pm$0.08 & 8 & 0.86\\					
\hline
\hline
\multicolumn{1}{c}{\rule{0pt}{2.5ex}  Isotope} &
\multicolumn{1}{c}{\rule{0pt}{2.5ex}  E$_{\mbox {\small proton}}$ [MeV]} &
\multicolumn{1}{c}{\rule{0pt}{2.5ex}  E$_{\gamma}$ [keV]} &
\multicolumn{1}{c}{\rule{0pt}{2.5ex}  t$_{1/2}$ [hour]} &
\multicolumn{1}{c}{\rule{0pt}{2.5ex}  d.o.f.} &
\multicolumn{1}{c}{\rule{0pt}{2.5ex}  $\chi^{2}_{red.}$} \\
\hline
\rule{0pt}{2.5ex}
$^{125}$Xe& 6.0 & 188.42 & 16.81$\pm$0.08 & 234 & 1.31 \\
          & 6.0 & 243.38 & 16.73$\pm$0.09 & 234 & 0.87\\
          & 6.5 & 188.42 & 16.84$\pm$0.33 & 27  & 1.13 \\
          & 6.5 & 243.38 & 16.97$\pm$0.45 & 27  & 1.62 \\ 
					& 7.0$^*$ & 188.42 & 17.27$\pm$0.31 & 150 & 1.09 \\
          & 7.0$^*$ & 243.38 & 17.36$\pm$0.32 & 150 & 0.89 \\
					& 7.0 & 188.42 & 16.87$\pm$0.13 & 115 & 0.80 \\
          & 7.0 & 243.38 & 17.17$\pm$0.22 & 115 & 1.20 \\
          & 7.4$^*$ & 188.42 & 16.71$\pm$0.11 & 107 & 0.96 \\
          & 7.4$^*$ & 243.38 & 17.05$\pm$0.19 & 107 & 0.99 \\
		      & 7.4 & 188.42 & 17.02$\pm$0.12 & 140 & 0.98 \\
          & 7.4 & 243.38 & 16.95$\pm$0.10 & 140 & 0.83 \\
          & 7.5$^*$ & 188.42 & 16.79$\pm$0.08 & 145 & 0.90 \\
          & 7.5$^*$ & 243.38 & 16.68$\pm$0.14 & 145 & 1.32 \\	
		      & 7.5 & 188.42 & 16.94$\pm$0.08 & 135 & 1.19 \\
          & 7.5 & 243.38 & 17.02$\pm$0.12 & 135 & 0.87 \\   																						
Error weighted mean&  & & 16.87$\pm$0.03 & 15 & 1.36\\		
\hline
\end{tabular}
\end{table}
\end{center}

\subsection{Systematic uncertainty of the half-life values}

The half-lives, measured in far geometry and in close geometry were in agreement within 0.09\% (for $^{125}$Cs) and 0.01\% (for $^{125}$Xe). Furthermore, selecting different background regions and using different methods for the net peak area determination led to a small, 0.2\% difference in the derived t$_{1/2}$ results. These values --- 0.09\% (for $^{125}$Cs), 0.01\% (for $^{125}$Xe) and 0.2\% were taken into account as systematic uncertainties. 

The new half-life value of $^{125}$Xe, derived in the present work is based on the counting of two $\gamma$-transitions. There is a small --- 0.33\% --- difference between the half-lives based on the 188.42 keV or the 243.38 keV lines. This difference may be explained by e.g. contributions from weak $\gamma$-background lines located close (within the resolution of the HPGe detector) to the peaks of our interest. The maximum of such a possible systematic uncertainty on the $^{125}$Cs half-life value was estimated in the following way. Using our HPGe detector it was not possible to resolve the 112 keV $\gamma$-peak of $^{125}$Cs from the 113.5 $\gamma$-peak of $^{125}$Xe. We estimated this contribution in each spectra by correcting the 243.38 keV peak areas using the known I$_{\gamma}$(113.50) / I$_{\gamma}$(243.38) branching ratios. After this subtraction the decay curve was fitted again, the resulted $^{125}$Cs half-lives were found to be in average 0.55\% higher then the ones based on the 526.47 keV peak. Therefore, this 0.55\% was considered as a systematic uncertainty.

\begin{center}
\begin{figure*}
\resizebox{1.0\columnwidth}{!}{\rotatebox{0}{\includegraphics[clip=]{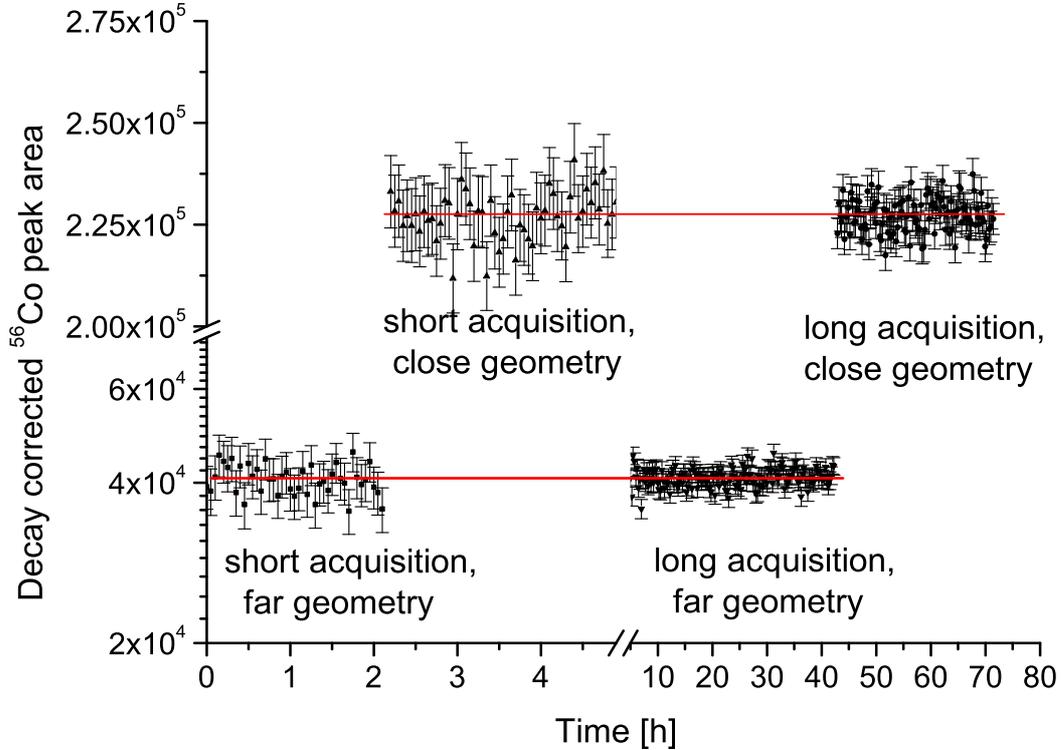}}}
\caption{\label{fig:Co_yield} Decay and deadtime corrected yield of the $^{56}$Co reference source used for checking the reliability of the deadtime determination and the stability of the counting setup efficiency both for the short and the long data acquisitions.
}
\end{figure*}
\end{center}

The reliability of the counting setup, namely, the effect of accidental detection efficiency changes and deadtime determination, can be studied using a long-lived reference source with precisely known half-life. The aluminum foil used to catch the $^{125}$Cs recoils, contained some iron impurity leading to the production of $^{56}$Co (t$_{1/2}$ = 77.2 d) through the $^{56}$Fe(p,n)$^{56}$Co reaction. This isotope emits several high intensity $\gamma$-lines (1037.840 keV, 1238.282 keV, 1360.215 keV, 1771.351 keV, 2015.181 keV and 2598.459 keV\footnote{The 846.77 keV peak of the $^{56}$Co is not resolved from the E$_{\gamma}$ = 846.51 keV line of $^{125}$Xe, therefore it was not used to monitor the stability of the counting setup.}) which could be used to monitor the stability of the counting setup. The deadtime correction provided by MAESTRO has been applied to the net peak areas of this source and the reliability of the deadtime was checked by studying the $^{56}$Co decay. Fig. \ref{fig:Co_yield} shows the decay and deadtime corrected 1238.282 keV yield of the $^{56}$Co reference source. The figure clearly shows that the decay and deadtime corrected yield is constant in time which proves the reliability of the deadtime values. The slope of the linear fit was always consistent with zero with maximum uncertainty of 0.2\% (which is in excellent agreement with a previous work \cite{Gyu12}) and this value can be considered as the upper limit of the systematic uncertainty characterizing the counting setup. 

The pulse pile-up effect can also influence the results of the $\gamma$-spectroscopy experiments \cite{Cla73,Gri07}. The initial counting rates characterizing the half-life determination of $^{125}$Cs, listed in Table \ref{tab:counting}, varied between 0.54 kHz and 1.33 kHz. No trend in the resulted half-life values as the function of the initial counting rates were observed. The counting rates characterizing the runs used to derive the $^{125}$Xe half-life were even lower, typically below 0.5 kHz. Therefore, it was concluded that event-loss due to pile-up does not influence the present data. Finally, it has to be noted that noble gasses tend to diffuse out of foils \cite{Ber03}, however, this effect was extensively studied in \cite{Hal16} and no loss was observed.

\section{Summary and conclusion}
\label{sec:results}

Table \ref{tab:res} lists the obtained half-life results for both the $^{125}$Cs and $^{125}$Xe isotopes. The uncertainties are statistical only and the $\chi^2_{red.}$ refers to the exponential fit. The values obtained from the different $\gamma$-countings are in good agreement. The values for the $^{125}$Cs and $^{125}$Xe half-lives and their statistical uncertainties are calculated as the weighted average of the measured samples. In the case of $^{125}$Xe the $\chi^2_{red.}$ value of the weighted average was higher than one, the uncertainty of the half-life has been multiplied by $\sqrt{\chi^2_{red.}}$. 
The total uncertainties are the quadratic sum of the statistical uncertainties and the above discussed systematic uncertainties. The final half-life results are t$_{1/2}$ = 44.35 $\pm$ 0.08 (stat.) $\pm$ 0.27 (syst.) minutes and t$_{1/2}$ = 16.87 $\pm$ 0.03 (stat.) $\pm$ 0.07 (syst.) hours, respectively. Since the new values are much more precise than any of the previous measurements, the result of our work is recommended as new adopted values. With the high precision decay half-lives of $^{125}$Cs and $^{125}$Xe determined in the present work, the systematic uncertainties for the $^{124}$Xe(p,$\gamma$)$^{125}$Cs cross section measurement can be reduced.

\section*{Acknowledgments}

This work was supported by NKFIH (K120666, NN128072) and by the \'UNKP-18-4-DE-449 New National Excellence Program of the Ministry of Human Capacities of Hungary. The authors thank M. Wiescher (University of Notre Dame, Department of Physics) for providing enriched $^{124}$Xe gas for the experiment. G. G. Kiss acknowledges support from the J\'anos Bolyai research fellowship of the Hungarian Academy of Sciences. This work was partially supported by the Scientific and Technological Research Council of Turkey (TUBITAK), Grants No.  114F487.

\end{document}